\documentclass[aps,twocolumn,prl]{revtex4}
\usepackage{graphicx}
\usepackage{amsfonts}
\usepackage{amssymb}
\usepackage{amsmath}

\newcommand{\vnm}{w_{n,m}}
\newcommand{\vum}{w_{n+1,m}}
\newcommand{\vdm}{w_{n-1,m}}
\newcommand{\dvnm}{\dot{w}_{n,m}}
\newcommand{\vnd}{w_{n,m-1}}
\newcommand{\vnu}{w_{n,m+1}}
\newcommand{\unm}{v_{n,m}}
\newcommand{\uum}{v_{n+1,m}}
\newcommand{\udm}{v_{n-1,m}}
\newcommand{\dunm}{\dot{v}_{n,m}}
\newcommand{\und}{v_{n,m-1}}
\newcommand{\unu}{v_{n,m+1}}

\begin{document}                                                                    
    
\title{The higher-dimensional
Ablowitz-Ladik model: from (non-)integrability and solitary waves to surprising collapse properties and more exotic solutions}
\author{
P. G.\ Kevrekidis$^1$, G.J. Herring$^2$, S. Lafortune$^3$ and Q.E. Hoq$^4$}
%R. Carretero-Gonz{\'a}lez$^2$
%and D. J. Frantzeskakis$^{3}$.
%} 
%\institute{\inst{1} 
\affiliation{
$^1$ Department of Mathematics and Statistics, University of Massachusetts, Amherst MA 01003-4515, USA \\
%}
%\author{G. Herring}
%\affiliation{
$^2$ Department of Mathematics and Statistics, Cameron 
University, Lawton, OK 73505, USA \\
%}
%\author{S. Lafortune}
%\affiliation{
$^3$ Department of Mathematics, College of Charleston,
Charleston, SC 29401, USA\\
$^4$ Department of Mathematics and Computer Science,
Western New England College, Springfield, MA, 01119, USA}
                                                                               
\begin{abstract} 
We propose a consideration of the properties of the 
two-dimensional Ablowitz-Ladik discretization of the ubiquitous
nonlinear Schr{\"o}dinger (NLS) model. We use singularity confinement
techniques to suggest that the relevant discretization should not be
integrable. More importantly, we identify the prototypical solitary
waves of the model and examine their stability, illustrating the
remarkable feature that near the continuum limit, this discretization
leads to the {\it absence of collapse} and {\it complete spectral wave 
stability},
in stark contrast to the standard discretization of the NLS. 
We also briefly touch upon the three-dimensional
case and generalizations of our considerations therein, and also
present some more exotic solutions of the model, such as 
exact line solitons and discrete vortices.
\end{abstract}

\maketitle

%\section{Introduction}
{\it Introduction}.
The nonlinear Schr{\"o}dinger (NLS) equation \cite{sulem,ablowitz1} 
is a prototypical 
dispersive nonlinear
partial differential equation (PDE) that has been central for
almost four decades now to a variety of areas in mathematical physics.
The relevant fields of
application vary from optics and propagation of the (envelope
of) the electric field in optical fibers \cite{hasegawa,kivshar},
to the self-focusing and 
collapse of Langmuir waves in plasma physics \cite{zakh2}
%zakh1
and from the behavior of deep water waves and 
freak waves 
in the ocean \cite{benjamin,onofrio} to the mean-field dynamics
of Bose-Einstein condensates in atomic physics 
\cite{stringari,pethick}.
%,emergent}. 

Much of the NLS literature and physical investigations
have been centered around the one-dimensional (1d) setting, due
not only to its mathematical and computational simplicity, but
also in major part due to its complete integrability via the
inverse scattering transform \cite{ablowitz1}. Yet, the higher
dimensional investigations of the NLS equation have important
elements of mathematics and physics in their own right, presenting
the possibility for self-focusing and wave collapse \cite{sulem}
that has remarkable manifestations in some of the physical areas 
represented above. It is, thus, not surprising that recent fundamental
investigations have been focused on related issues, such
as e.g. the observation of the self-similarly collapsing
solitary wave of the two-dimensional (2d) NLS equation
(the so-called Townes soliton) \cite{fibich}, and on how to avoid the
relevant collapse phenomena by means of temporal \cite{ueda,us_exp}
or spatial \cite{fibich2} variations of the nonlinearity, or by 
posing the problem on a lattice \cite{arevalo}.

In the present work, we revisit the higher
dimensional NLS equation and consider its properties
on a spatial lattice. This is an interesting approach in its own
right, at least in part because many of the problems associated
with the optics (namely, optical waveguides \cite{review_opt}) or 
atomic physics (namely, Bose-Einstein condensates --BECs-- in optical
lattices \cite{mplb}) are inherently associated with such 
discrete models. Another relevant motivation is that of computation,
since even computing with the continuum model proper requires posing 
the problem on a computational grid (with a finite, but small spacing).
However, in our investigations herein, we will not use the ``canonical''
discrete form of the NLS equation, the so-called DNLS model 
\cite{IJMPB,dnls_book}. Instead, we will consider
an unconventional discretization that has yielded considerable
insight (although, for a quite different reason, as illustrated
below), namely the Ablowitz-Ladik (AL-NLS) model \cite{ablowitz1}, which in its two-dimensional
form reads:
\begin{eqnarray}
i\dot u_{n,m}  =  &-& \varepsilon \left( {u_{n + 1,m}  + u_{n - 1,m}  + u_{n,m + 1}  + u_{n,m - 1}  - 4u_{n,m} } \right) 
\nonumber
\\
&+& \frac{\sigma }{4}\left| u \right|^2 \left( {u_{n + 1,m}  + u_{n - 1,m}  + u_{n,m + 1}  + u_{n,m - 1} } \right)
\label{ALEqn}
\end{eqnarray}
In particular, the fundamental difference from the DNLS here is that
in addition to the centered-difference approximation of the 
Laplacian (with $\varepsilon=1/\Delta x^2$ being the coupling strength), 
a nearest-neighbor average is used to discretize
the cubic nonlinearity of the continuum limit $ \sigma |u|^2 u$
(instead of a local term $\sigma |u_{n,m}|^2 u_{n,m}$ in the DNLS). 
The cubic nonlinearity physically represents the 
Kerr effect in optics (i.e., the dependence of the refractive
index of the optical material on the light intensity) or the 
mean-field approximation of interatomic interactions as a nonlinear
self-action in 
%the case of alkali vapors in 
BECs.

Our main findings and presentation are as follows. In the next
section, we use the technique of singularity confinement to illustrate
that, by analogy to the continuum model, the 
AL-NLS model is unlikely to be integrable in two spatial dimensions.
We then embark on a systematic analysis of the model's solitary 
waves and their stability and substantiate a surprising result,
namely that the AL-NLS discretization possesses {\it spectrally stable}
solitons for arbitrarily small lattice spacings $\Delta x$.
We then consider the generalization of this result in the 3d setting.
Lastly, we present some more exotic solutions of the AL-NLS model,
such as the analytically exact (but unstable) line soliton and
the $x$-shaped discrete vortex.

{\it Singularity Confinement.} To determine whether the AL-NLS is completely integrable, 
we use an integrability detector designed for difference or differential-difference equations, 
namely the singularity confinement (SC) criterion \cite{SC,Gram}.
The SC deals with the spontaneous appearance
of a singularity at some point in the lattice. The criterion is satisfied
if the singular behavior is confined in a finite region of the lattice. 
In order to apply the SC criterion,
we re-interpret the AL-NLS 
as an iteration for the real and imaginary parts of $u=v+iw$: 
\begin{equation}
\nonumber
\begin{aligned}
&\hspace{-.2cm}\uum=\frac{4\dvnm+16\epsilon\unm}{4\epsilon-\sigma |u|^2}-\unu-\udm-\und,&\\
&\hspace{-.2cm}\vum=\frac{-4\dunm+16\epsilon\vnm}{4\epsilon-\sigma |u|^2}-\vnu-\vdm-\vnd.&
\label{AL2ds}
\end{aligned}
\end{equation}
If, at some point $(n_0,m_0)$, 
the denominator above is zero
%the pair $\unm$, $\vnm$ 
%is such that $4\epsilon-\sigma |u|^2$ is zero 
for some time $t=t_0$, then the iterates  $\uum$, $\vum$ will be singular. 
%at $t=t_0$. 
We thus set
 $4\epsilon-\sigma |u_{n_0,m_0}|^2=(t-t_0)\alpha(t)$ (for an arbitrary function $\alpha$ nonzero at $t=t_0$), 
 and let $\unm$ and $\vnm$ for $n=n_0$ (with $m\neq m_0$) and $n=n_0-1$ to 
 be regular and arbitrary.
Computation of the iterates indicates that the singular behavior at $t=t_0$ is not confined but rather propagates indefinitely for larger values of $n$. 
This result implies that the AL-NLS is not 
completely integrable.

{\it Solitary Waves and Stability}. We start by recalling that
the continuum analog of the 2d model is unstable (due to an 
instability which is weaker than exponential \cite{comech}) towards
self-similar collapse \cite{sulem}. The solitonic standing wave
continuum solution (in the form $u(x,y,t)=\exp(i \Lambda t) v(x,y)$,
taking
$\Delta x \rightarrow 0$ and $\sigma=-1$ in Eq. (\ref{ALEqn}))
is the so-called Townes soliton \cite{townes} with a squared $L^2$-norm
(i.e., optical power or number of atoms in BEC) equal to $P_c \approx 11.7$.
This is the separatrix between the collapse regime, arising for
 powers $P>P_c$, and dispersion occurring for $P<P_c$.

\begin{figure}[tbp]
\includegraphics[width=2.5cm,height=5cm,angle=0,clip]{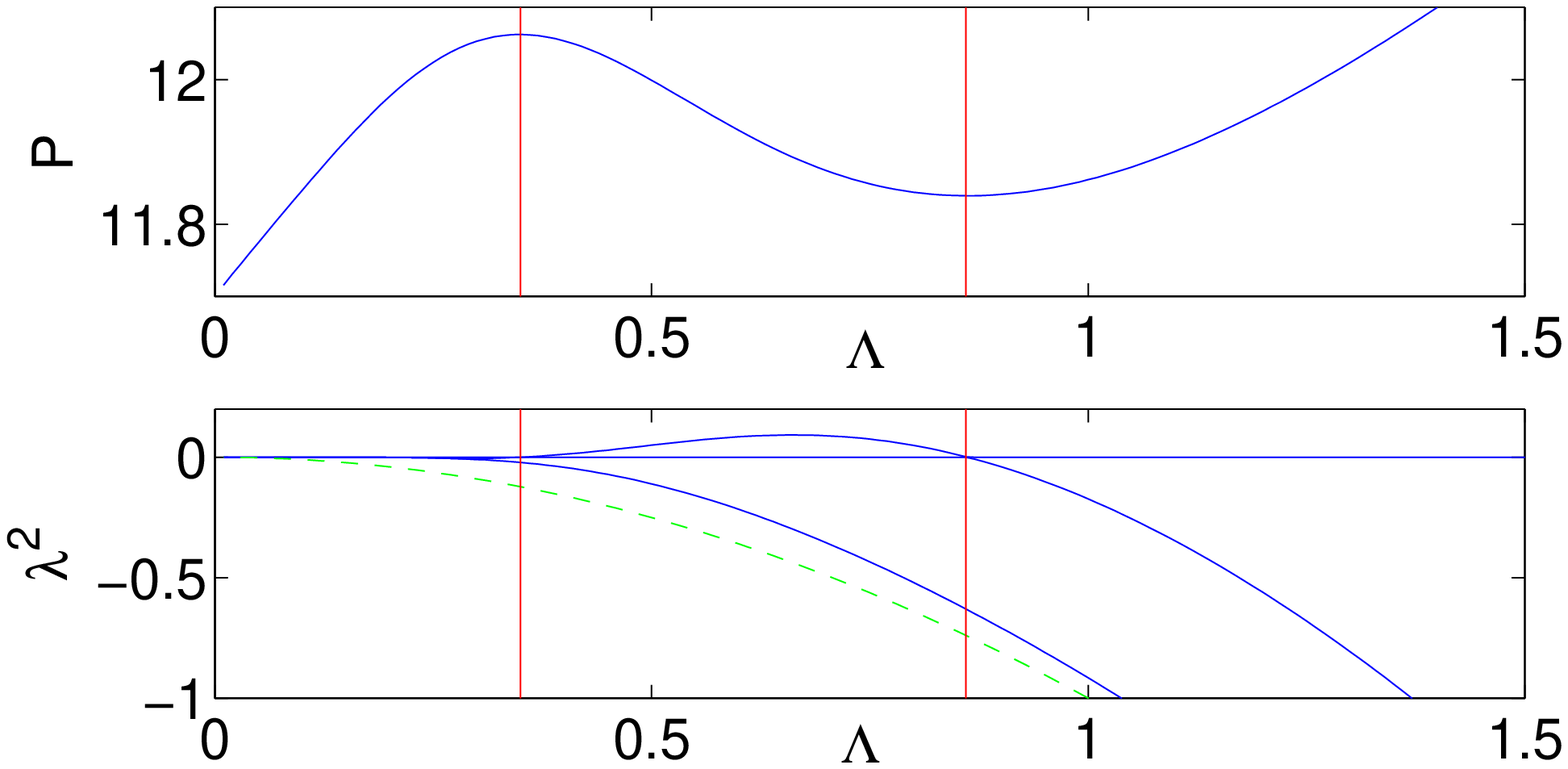}
\includegraphics[width=2.5cm,height=5cm,angle=0,clip]{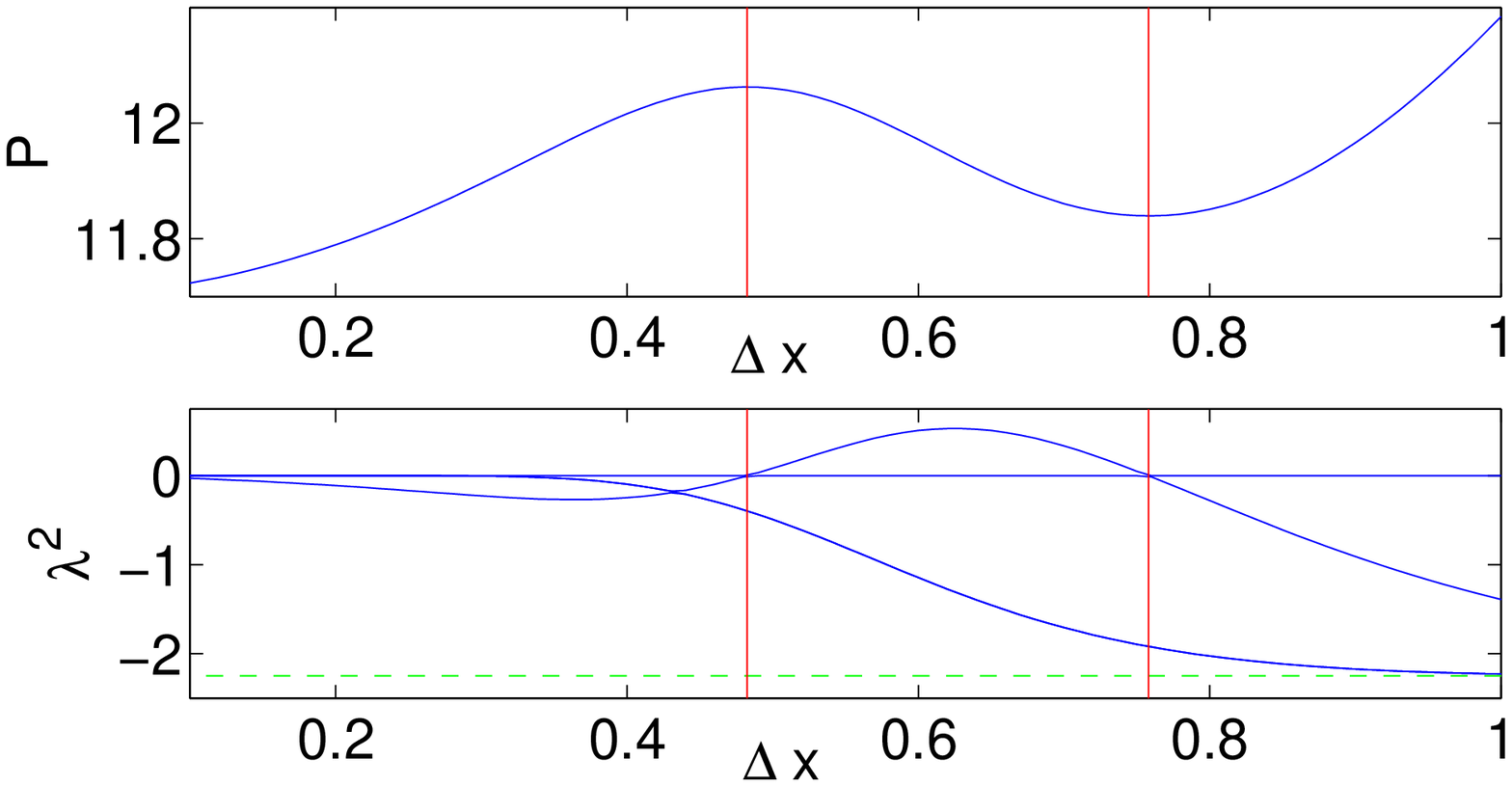}
\includegraphics[width=2.5cm,height=5cm,angle=0,clip]{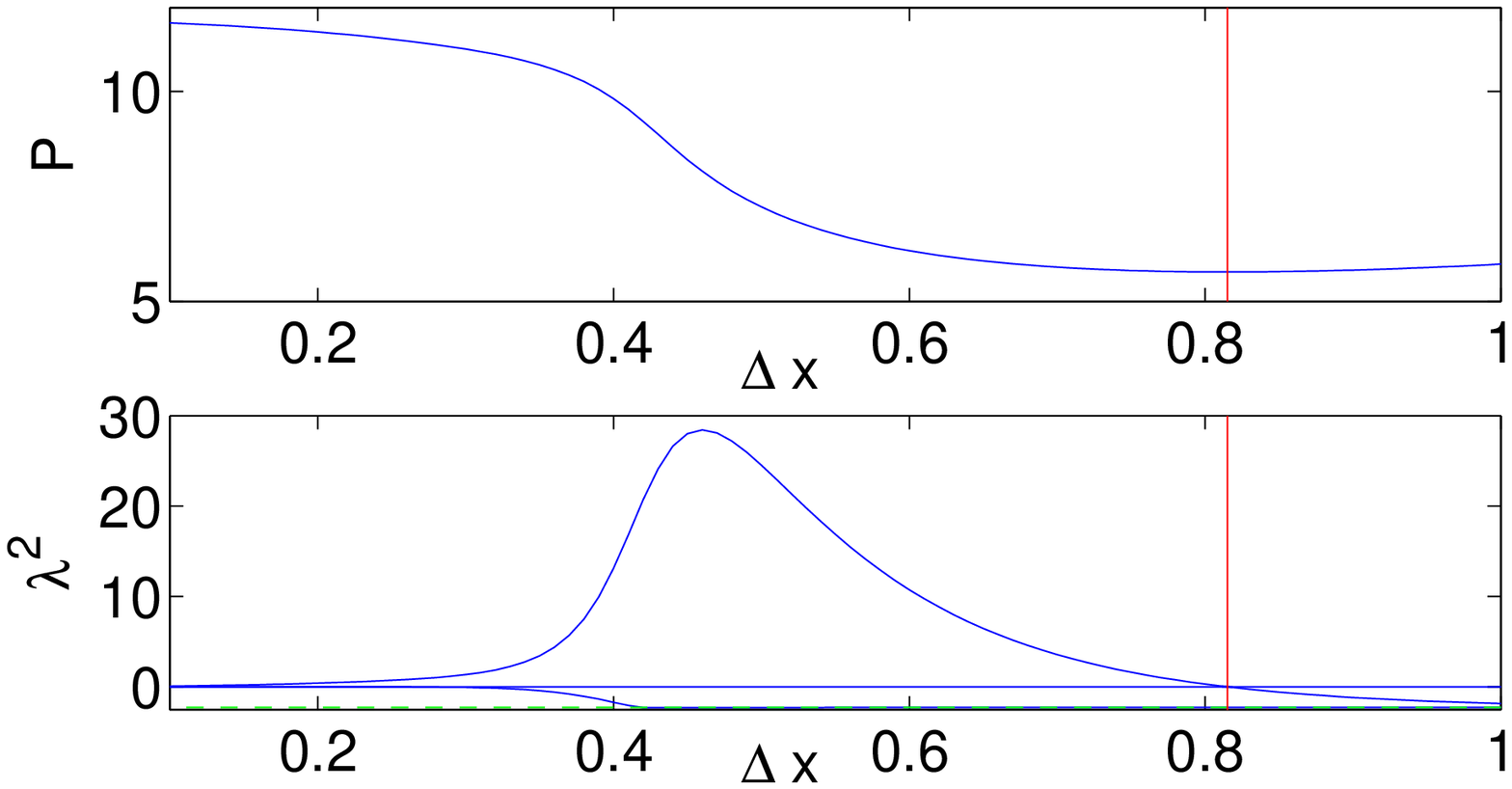}
\caption{(Left) Dependence of the AL-NLS fundamental 
soliton power P (Eq. (\ref{EnergyEqn}))
and of the square eigenvalues of the linearization $\lambda^2$
on the wave frequency $\Lambda$ ($\Delta x=1$). Vertical red lines bound the
instability region of $d P/d \Lambda<0$. Green dashed line denotes
the (lower) edge of the phonon band. (Middle) Same but for the 
variation as a function of the grid spacing $\Delta x$ ($\Lambda=1.5$). 
(Right) Same as the middle panel, but now for the regular DNLS
discretization.} 
\label{fig1}
\end{figure}

In the discrete case of the AL-NLS, we also seek standing waves
in the form $u_{n,m}(t)=\exp(i \Lambda t) v_{n,m}$. Fixing 
$\sigma=-1$ (i.e., focusing nonlinearity), 
we can seek such solutions either as a function of
$\Lambda$ or (equivalently, up to a simple rescaling) as a function
of $\Delta x$. We present both cases in Fig. \ref{fig1}. The continuum limit is obtained as $\Lambda$ or, respectively,
$\Delta x \rightarrow 0$. The former case has a particularity with
respect to stability (which is part of the reason as to why it
is presented herein). In particular, the so-called Vakhitov-Kolokolov
criterion suggests that for {\it fundamental} solutions 
\cite{sulem,IJMPB,dnls_book}, the soliton stability is {\it solely} determined
by the sign of the quantity $d P/d \Lambda$. In the AL-NLS case,
interestingly (a direct generalization of the corresponding 1d
conservation law \cite{ablowitz1}), the power is defined as:
\begin{equation}
P = \sum -\frac{\sigma}{4}{\ln \left( {1 - \frac{\sigma }{{4\varepsilon }}\left| u \right|^2 } \right)}, 
\label{EnergyEqn}
\end{equation}
such that it gives the continuum power in the limit. When, in our
units, this quantity is positive, the solitary wave should be stable,
while its negativity should imply exponential instability. 

\begin{figure}[tbp]
\includegraphics[width=4cm,height=5cm,angle=0,clip]{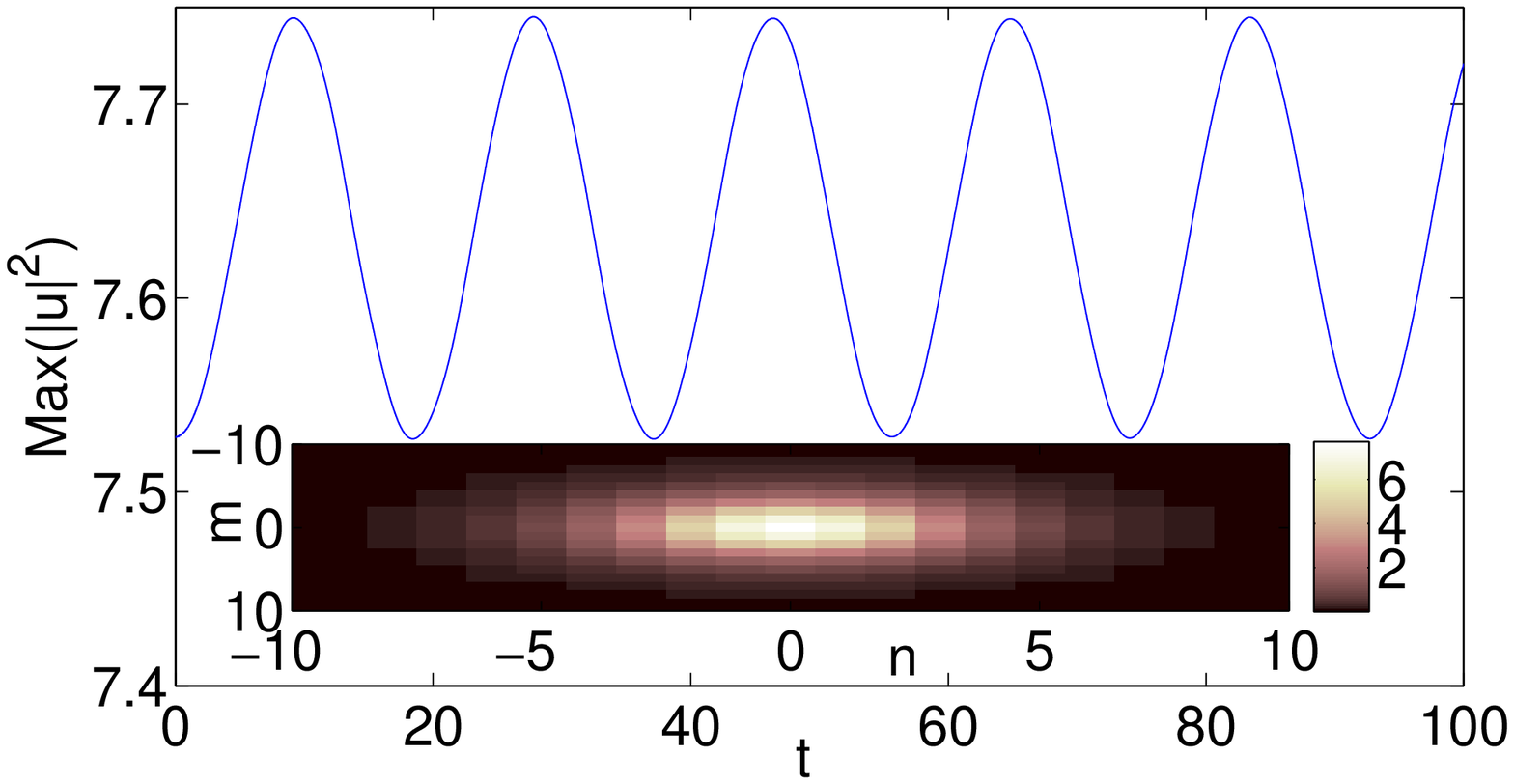}
\includegraphics[width=4cm,height=5cm,angle=0,clip]{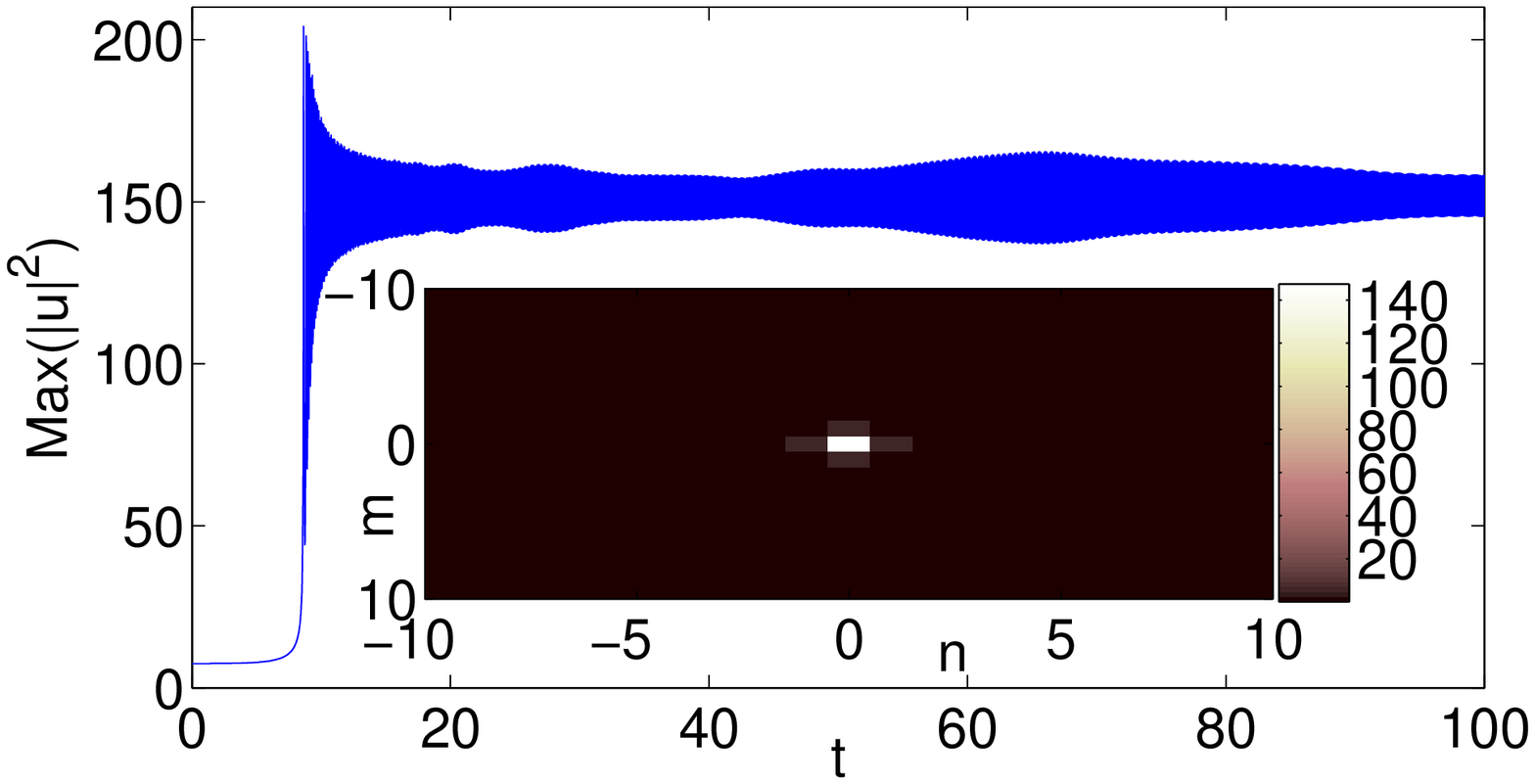}
\includegraphics[width=4cm,height=5cm,angle=0,clip]{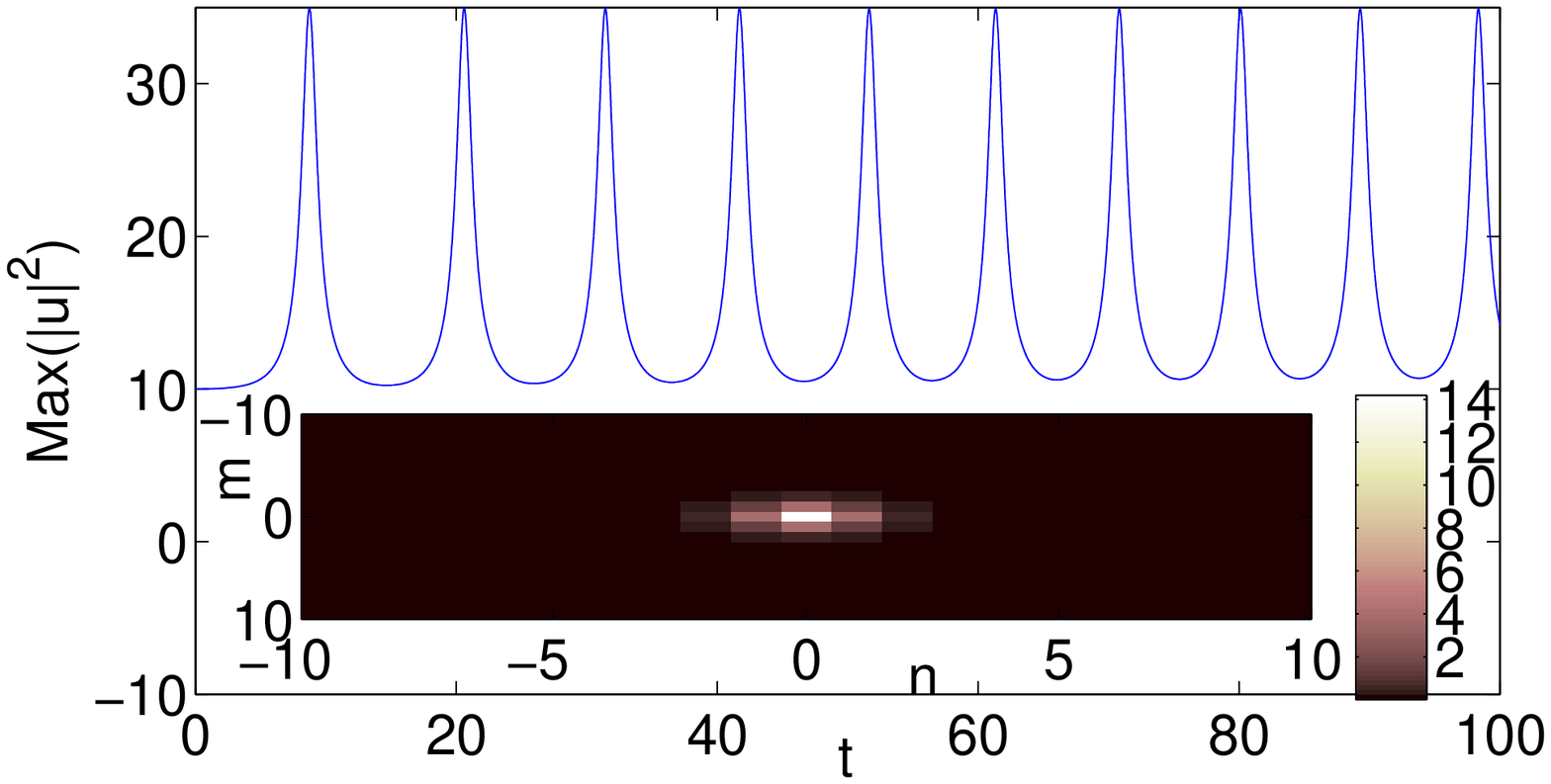}
\includegraphics[width=4cm,height=5cm,angle=0,clip]{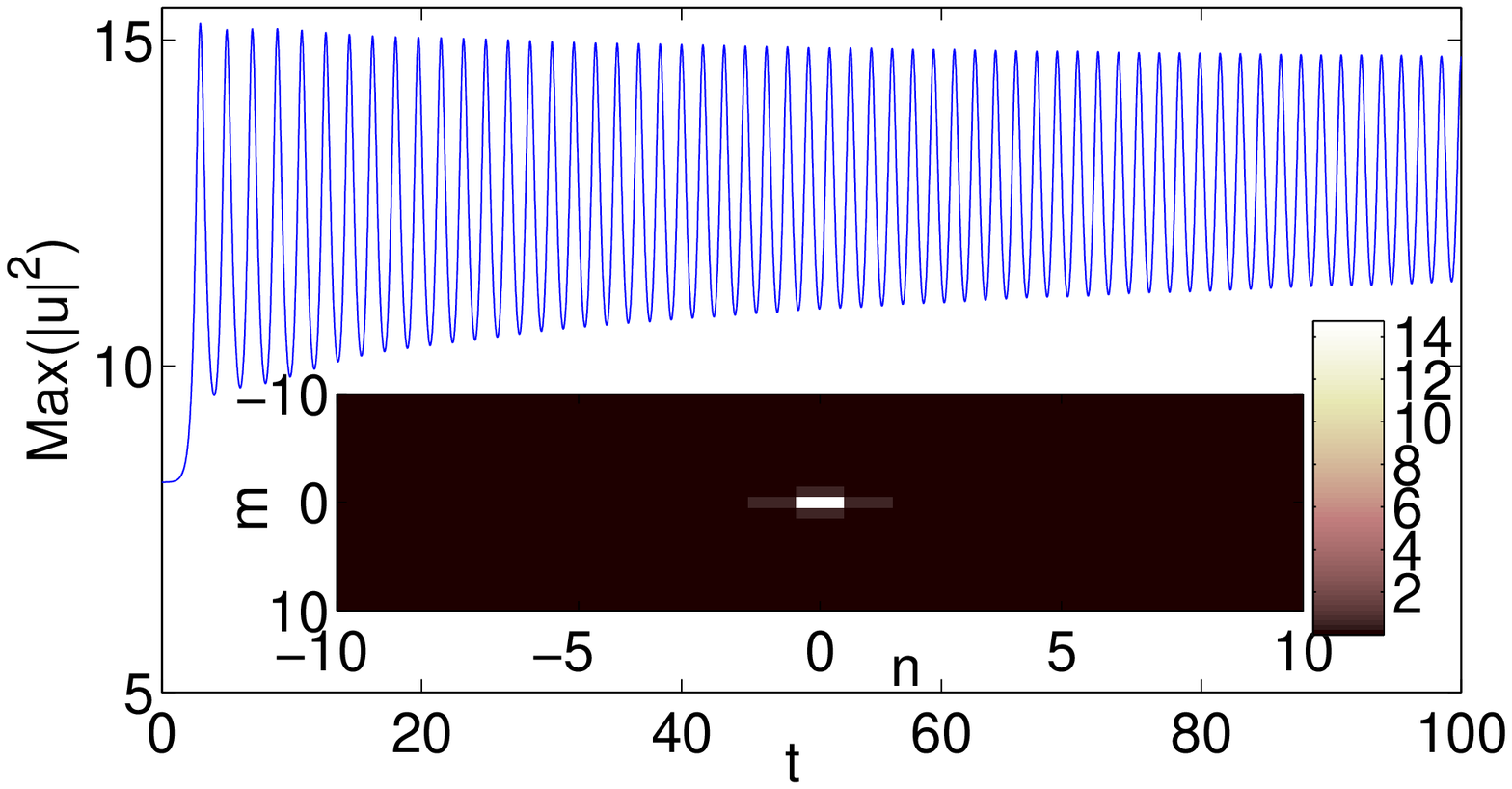}
\caption{Dynamical Evolution of the soliton solution maximum for AL-NLS
(left) and DNLS (right) for $\Delta x=0.2$ (top) and $0.6$ (bottom).
The insets show the contour plot of $|u_{n,m}(100)|^2$. Notice the
stability of the AL-NLS for $\Delta x=0.2$ (all other cases lead to collapse)
} 
\label{fig2}
\end{figure}

It is perhaps worthwhile to commence our comparison of the 
discrete solitary wave results to the corresponding continuum
model through the intensely studied DNLS model \cite{IJMPB,dnls_book}.
There it is known (see also the right panel of Fig. \ref{fig1}
for the relevant properties as a function of $\Delta x$) that as
the continuum limit is approached, at the critical threshold of
$\varepsilon=\Lambda$ (for spacings or frequencies below that),
the solitary waves become {\it exponentially unstable}. 
This 
dynamical instability, manifested the right panels of Fig. \ref{fig2} for two
different values of the spacing $\Delta x$, is connected to
the collapse of the continuum model and consists of a ``quasi-collapse''
phenomenon, whereby all the power of the solution is ``collected'' on
a single site; the discrete model with its discrete power conservation
and its spatial grid scale disallows a true self-similar collapse
leading, in principle, to a finite time singularity. It should be noted
that once unstable, the fundamental soliton of the 
DNLS model remains unstable throughout the
parametric continuation to the continuum limit. The spectral picture
of the linearization around the solitary wave (to examine the fate
of small perturbations) reveals that in addition to a pair of 
eigenvalues always at the origin [due to the U$(1)$ phase invariance
of the model], there are 6 other pairs near the origin, approaching
it, as the 2d continuum limit draws near. Of these, two identical
pairs are associated with the translations along the x- and y-directions
(whose corresponding invariances are restored in the limit) and which
always remain stable (i.e., with $\lambda^2<0$). On the contrary, the
eigenvalue pair with $\lambda^2>0$ connected with the slope condition
and with the quasi-collapse (focusing instability in the language
of \cite{sivan}) is also connected with the additional symmetry of
the 2d limit, namely the so-called pseudo-conformal invariance \cite{sulem},
which allows the self-similar reshaping of the solution (and, hence,
the access to the collapse dynamics of the continuum limit).

One of the remarkable findings of the present work is that AL-NLS
model are the {\it fundamentally} different spectral properties
of its solitary wave. In particular, as can be seen in the top
panel of Fig. \ref{fig1}, the change of sign of 
$d P/d \Lambda$ (and, accordingly, the instability) 
arises for $0.34 < \Lambda < 0.87$ for
$\Delta x=1$, or, respectively, the instability emerges for 
$0.4825<\Delta x< 0.758$ (for $\Lambda=1.5$). {\it However},
as the continuum limit is approached either through $\Lambda \rightarrow 0$,
or through $\Delta x \rightarrow 0$, the relevant eigenvalue associated
(in the limit) with the pseudo-conformal invariance has $\lambda^2 < 0$.
Furthermore, the squared eigenvalues 
double pair associated with translations also approaches zero
from the negative side. 
This implies that arbitrarily close to the limit, the AL-NLS discretization
offers an alternative of NLS {\it free of collapse instabilities}. 
This feature is also evident in the case of the dynamical evolution
of Fig. \ref{fig2}, whose top left panel illustrates that for small values
of $\Delta x$, the solitary waveform is not subject to collapse
(as it may be e.g. within the instability interval
arising for higher $\Delta x$, as illustrated in the
figure for $\Delta x=0.6$).

Further consideration of this feature suggests that it is a
{\it particular} trait of {\it critical} settings, which are
at the very special (yet, physically realizable and important)
separatrix between subcritical settings where the solitary waves
are dynamically stable and supercritical ones, where the waves
are exponentially unstable. In this critical case, the linear
spectrum possesses an additional zero
eigenvalue pair (associated with the pseudo-conformal invariance),
which permits the reshaping of the solution under
the action of the group of rescalings, and hence paves the way
for the emergence of self-similar collapse. Discreteness
can then shift this pair along the imaginary axis or along
the real axis (as it is well documented to potentially do with
respect to translational eigenvalues also \cite{IJMPB,dnls_book}).
The AL-NLS discretization turns out to be a prototypical example
whereby the eigenvalue pair formerly associated with the pseudo-conformal
invariance is perturbed in a stable way (moves along the imaginary
axis of the spectral plane), upon discretization and hence, this
model allows infinitesimally small spacings to give rise to 
{\it collapse-free} dynamics. 

%\begin{figure}[tbp]
%\includegraphics[width=2.5cm,height=5cm,angle=0,clip]{al_fig1_new.eps}
%\includegraphics[width=2.5cm,height=5cm,angle=0,clip]{al_fig1a.eps}
%\includegraphics[width=2.5cm,height=5cm,angle=0,clip]{al_fig_dnls1.eps}
%%\includegraphics[width=4cm,height=5cm,angle=0,clip]{al_fig2.eps}
%\caption{(Left) Dependence of the AL-NLS fundamental 
%soliton power P (Eq. (\ref{EnergyEqn}))
%and of the square eigenvalues of the linearization $\lambda^2$
%on the wave frequency $\Lambda$ ($\Delta x=1$). Vertical red lines bound the
%instability region of $d P/d \Lambda<0$. Green dashed line denotes
%the (lower) edge of the phonon band. (Middle) Same but for the 
%variation as a function of the grid spacing $\Delta x$ ($\Lambda=1.5$). 
%(Right) Same as the middle panel, but now for the regular DNLS
%discretization.} 
%\label{fig1}
%\end{figure}

%\begin{figure}[tbp]
%\includegraphics[width=4cm,height=5cm,angle=0,clip]{al_dx_02.eps}
%\includegraphics[width=4cm,height=5cm,angle=0,clip]{dnls_dx_02.eps}
%\includegraphics[width=4cm,height=5cm,angle=0,clip]{al_dx_06.eps}
%\includegraphics[width=4cm,height=5cm,angle=0,clip]{dnls_dx_06.eps}
%\caption{Dynamical Evolution of the soliton solution maximum for AL-NLS
%(left) and DNLS (right) for $\Delta x=0.2$ (top) and $0.6$ (bottom).
%The insets show the contour plot of $|u_{n,m}(100)|^2$. Notice the
%stability of the AL-NLS for $\Delta x=0.2$ (all other cases lead to collapse)
%} 
%\label{fig2}
%\end{figure}

\begin{figure}[tbp]
\includegraphics[width=5cm,height=5cm,angle=0,clip]{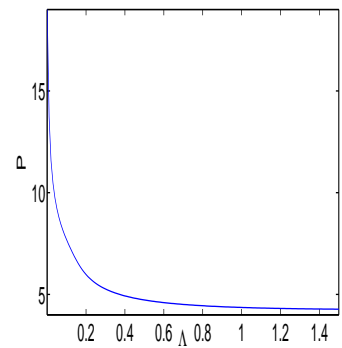}
\caption{Power vs. frequency plot for the unstable $dP/d\Lambda<0$ 3d AL-NLS
case ($\Delta x=1$).
} 
\label{fig3}
\end{figure}

{\it Higher-dimensional Generalizations and More Complex Waveforms}.
The above discussion paves the way for understanding
the approach to the continuum limit of the 3d solitary wave in the
3d AL-NLS generalization. The relevant power-frequency diagram is
shown in Fig. \ref{fig3}, clearly illustrating through its negative
slope the completely unstable approach to the limit
(since $dP/d \Lambda < 0$). In this case, the continuum limit has
$P \approx 18.82$ \cite{yaron} and the soliton is {\it strongly} 
(exponentially) unstable due to the supercritical collapse. Hence,
the discretization has no choice but to respect the relevant limit
and thus infinitesimally small grid spacings are unable to provide
stabilization {\it irrespectively} of the specific form of
the discrete model. 

\begin{figure}[tbp]
\includegraphics[width=4cm,height=5cm,angle=0,clip]{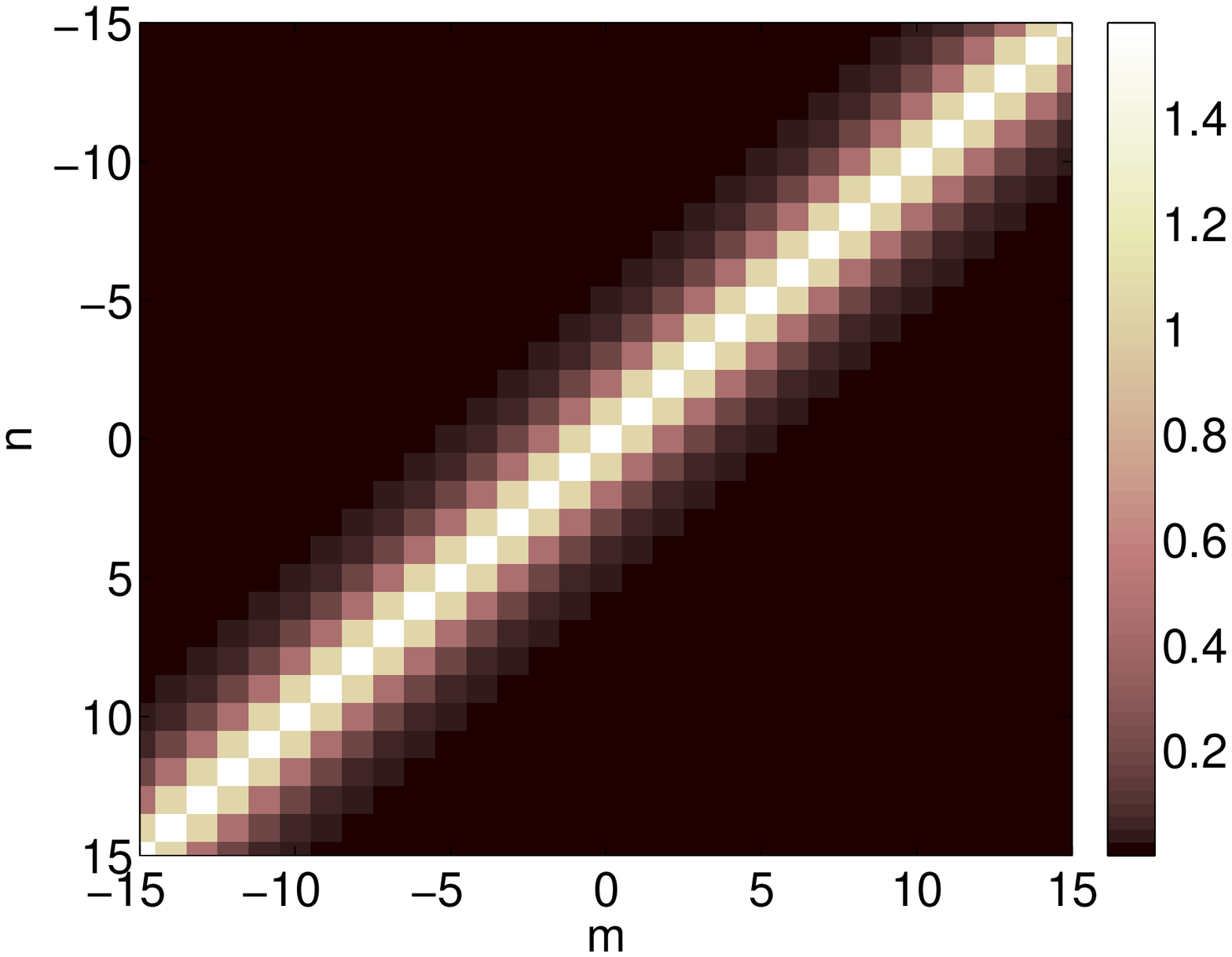}
\includegraphics[width=4cm,height=5cm,angle=0,clip]{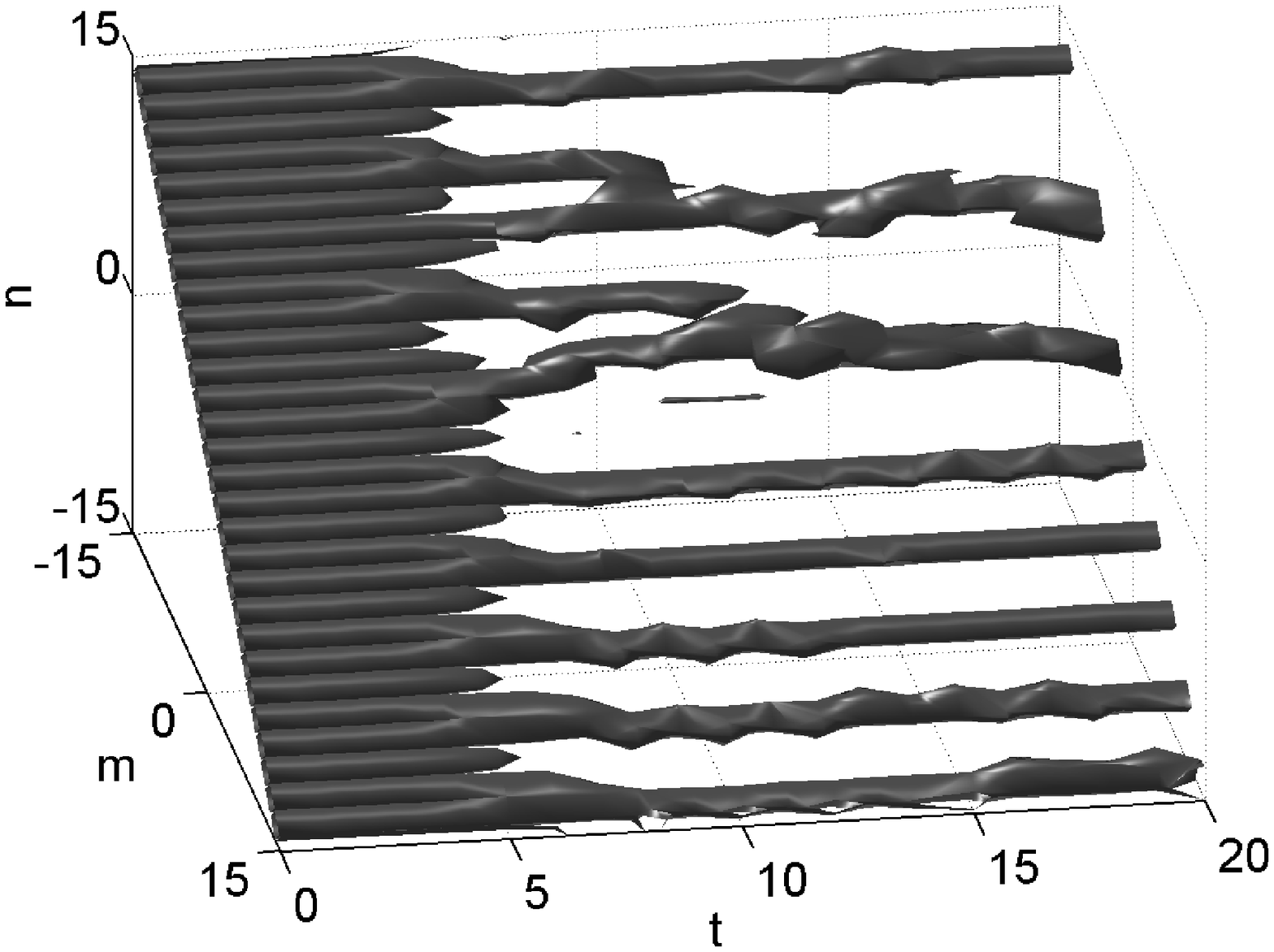}
\includegraphics[width=4cm,height=5cm,angle=0,clip]{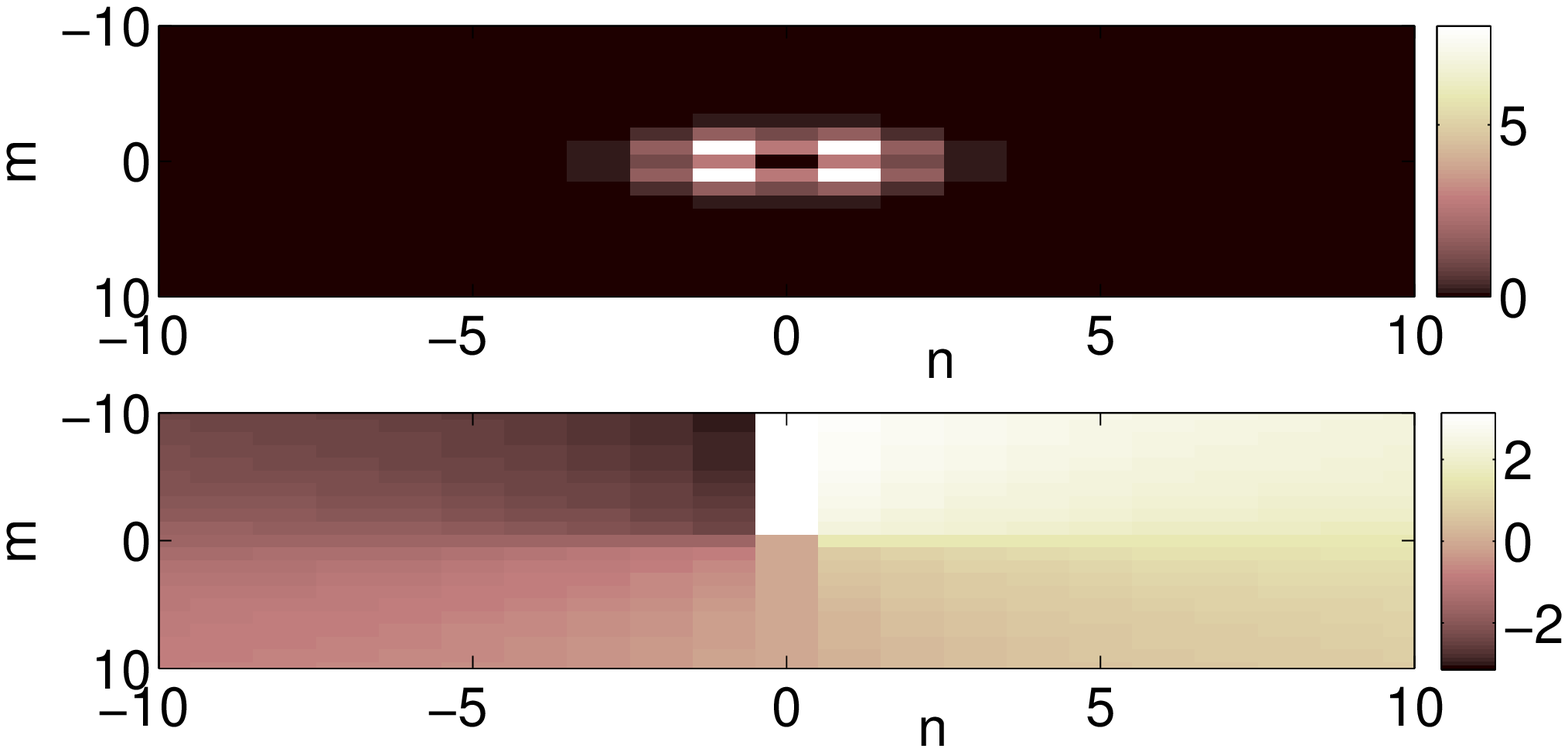}
\includegraphics[width=4cm,height=5cm,angle=0,clip]{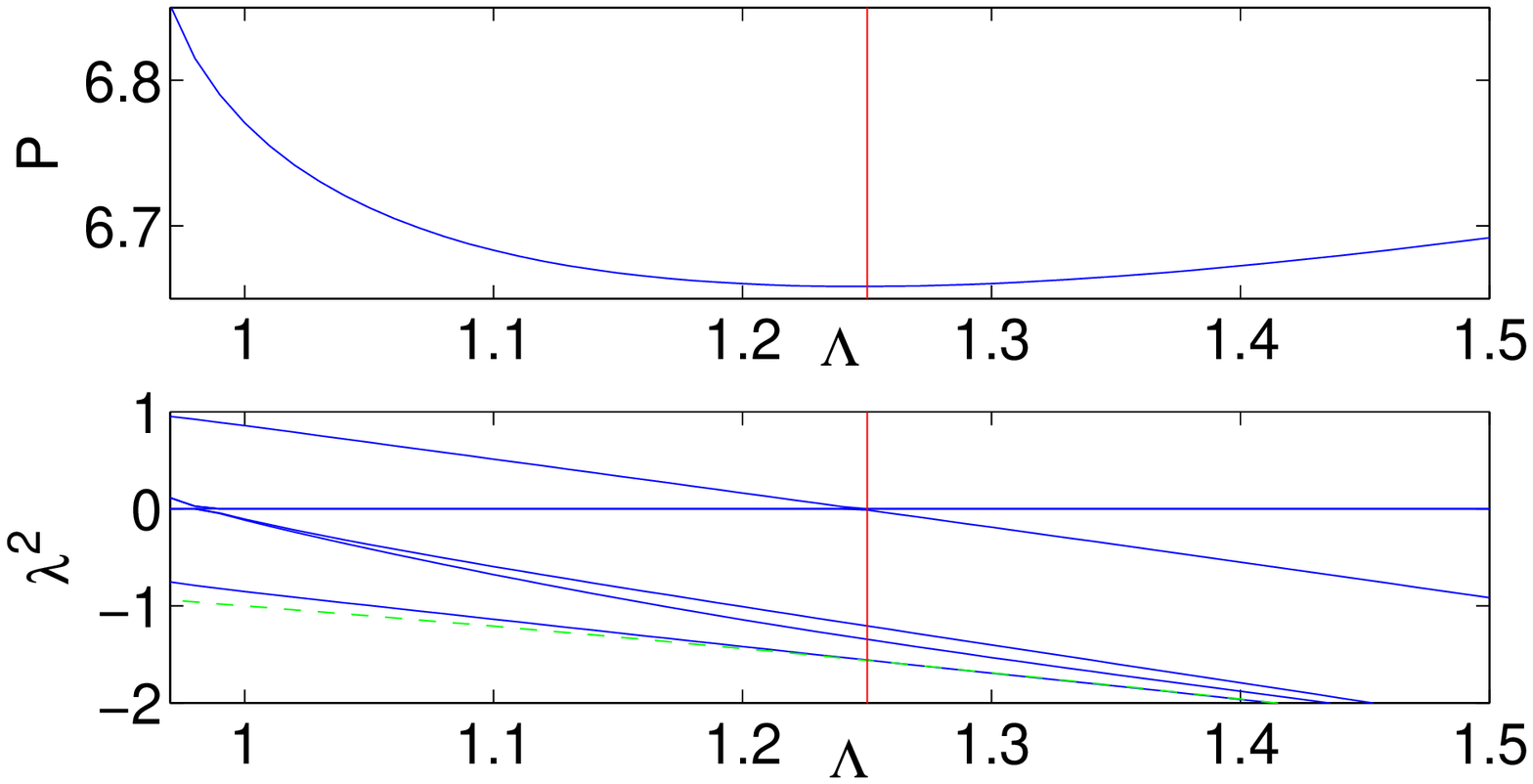}
\caption{The top panels show the exact line soliton contour (left) and
  its dynamical
instability evolution (right); $\Lambda=2 \varepsilon=1$. The bottom ones show the vortex square modulus
and phase for $\Lambda=1.1$ 
(left) and its power and linearization spectrum similar to
Fig. \ref{fig1} (right).
} 
\label{fig4}
\end{figure}

Lastly, we consider some more complex waveforms that
can arise in the AL-NLS model.
%On the one hand, 
It is possible 
to obtain explicit solutions in the form of quasi-1d solitons
e.g., 
\begin{eqnarray}
u_{n,m}=A {\rm sech}(\alpha n + \beta m + x_0) \exp(i \Lambda t)
\label{eqn3}
\end{eqnarray}
where the parameters satisfy $A=\pm \sqrt{(\Lambda^2+8 \Lambda 
\varepsilon)/(4 \varepsilon)}$, and 
$\alpha=\beta=\cosh^{-1}[(\Lambda+4 \varepsilon)/(4 \varepsilon)]$.
Such a solution is depicted in the top left panel of Fig. \ref{fig4},
but the linearization around it illustrates that it is highly unstable
and its dynamics spontaneously lead to filamentation and the formation
of localized solitary waves of the type considered previously, as seen
in the top right panel of Fig. \ref{fig4}. On the other hand, there
also exist more complex solutions, such as the discrete $x$-shaped vortices
of the bottom left panel of Fig. \ref{fig4}. Such solutions exist
in the DNLS equation, because of its anti-continuum limit $\varepsilon=0$
\cite{peli2d} but do not persist in the continuum limit, hence their
existence is not guaranteed in the AL-NLS model. Nevertheless, we find
here that they exist and are quite robust, becoming unstable due
to a real eigenvalue pair for $\Delta x<1.25$ (bottom right panel of
Fig. \ref{fig4}). Additional eigenvalue
pairs emerge for $\Delta x<0.985$. It is interesting to note that the
corresponding $x$-shaped vortex in the DNLS case becomes unstable due
to complex eigenvalue quartets for  
spacing values $\Delta x < 1.178$.

{\it Conclusions}. In this study, we considered the Ablowitz-Ladik
discretization of the NLS in higher
dimensional settings. We illustrated, via singularity confinement, 
that the model is unlikely to be integrable, yet,
due to the critical nature of the nonlinearity in 2d, it possesses
some remarkable features, including robust and spectrally stable
solitary waves even infinitesimally close to the continuum limit,
explicit analytical solutions and more complex vortex waveforms.
This study raises many fundamental issues.
It would be relevant to obtain a systematic understanding of how 
different NLS discretizations in the critical case affect the 
pseudo-conformal invariance (and whether there might conceivably
exist one that preserves the symmetry). It would
also be useful to examine how different types of discretizations
affect other classes of critical models, such as the critical
generalized KdV \cite{bona}. Lastly, it would be relevant
to obtain a systematic classification (analogous to the one existing
in DNLS \cite{peli2d}) of the solutions of the AL-NLS and their
stability properties, in two- and higher dimensions.


\begin{thebibliography}{99}

\bibitem{sulem} C. Sulem and P.L. Sulem, 
\newblock  {\it The Nonlinear Schr{\"o}dinger Equation}, 
Springer-Verlag (New York, 1999).

\bibitem{ablowitz1} M.J. Ablowitz, B. Prinari and A.D. Trubatch,
{\it Discrete and Continuous Nonlinear Schr{\"o}dinger Systems},
Cambridge University Press (Cambridge, 2004).

\bibitem{hasegawa} A. Hasegawa, {\it Solitons in Optical
Communications}, Clarendon Press (Oxford, NY 1995).

%\bibitem{malomed} B.A. Malomed, 
%{\it Variational methods in nonlinear fiber optics and related fields}, 
%Progress in Optics {\bf 43}, (2002) 69-191.

\bibitem{kivshar} Yu.S. Kivshar and G.P. Agrawal,
{\it Optical solitons: from fibers to photonic crystals},
Academic Press (San Diego, 2003).


%\bibitem{zakh1} V.E. Zakharov,
%{\it Collapse and Self-focusing of Langmuir Waves},
%\newblock Handbook of Plasma Physics, (M.N. Rosenbluth and R.Z. Sagdeev
%eds.), vol. 2 (A.A. Galeev and R.N. Sudan eds.), 81-121, Elsevier (1984).

\bibitem{zakh2} V.E. Zakharov,
%{\it Collapse of Langmuir waves}, 
Sov. Phys. JETP {\bf 35} (1972) 908-914.

\bibitem{benjamin} T.B. Benjamin and J.E. Feir,
%{\it The disintegration of wavetrains in deep
%water, Part 1}, 
J. Fluid Mech. {\bf 27}, (1967) 417-430.

\bibitem{onofrio} M. Onorato {\it et al.}, 
%A.R. Osborne, M. Serio, and S. Bertone, 
%{\it Freak waves in random oceanic sea states},
                      Phys. Rev. Lett. {\bf 86} (2001) 5831-5834.

\bibitem{stringari} L.P. Pitaevskii and S. Stringari,
{\it Bose-Einstein Condensation}, Oxford University Press (Oxford, 2003).

\bibitem{pethick} C.J. Pethick and H. Smith,
{\it Bose-Einstein condensation in dilute gases}, Cambridge University
Press (Cambridge, 2002).

%\bibitem{emergent} P.G. Kevrekidis, D.J. Frantzeskakis and R.
%Carretero-Gonz{\'a}lez, {\it Emergent Nonlinear Phenomena in
%Bose-Einstein Condensates}, Springer-Verlag (Berlin, 2008).

\bibitem{fibich} K.D. Moll {\it et al.}, 
%A.L. Gaeta and G. Fibich,
Phys. Rev. Lett. {\bf 90}, 203902 (2003).

\bibitem{ueda}  H. Saito and M. Ueda,
Phys. Rev. Lett. {\bf 90}, 040403 (2003);
F.Kh. Abdullaev {\it et al.}, 
%J.G. Caputo, R.A. Kraenkel and B.A. Malomed,
Phys. Rev. A {\bf 67}, 013605 (2003).

\bibitem{us_exp} M. Centurion {\it et al.}, 
%M.A. Porter, P.G. Kevrekidis
%and D. Psaltis, 
Phys. Rev. Lett. {\bf 97}, 033903 (2006).

\bibitem{fibich2} Y. Sivan {\it et al.}, 
%G. Fibich and M.I. Weinstein,
Phys. Rev. Lett. {\bf 97}, 193902 (2006).

\bibitem{arevalo} E. Ar{\'e}valo, Phys. Rev. Lett. {\bf 102},
224102 (2009).

\bibitem{review_opt} D. N.\ Christodoulides,
F.\ Lederer, and Y.\ Silberberg,
Nature \textbf{424}, 817 (2003); A. A.\ Sukhorukov
{\it et al.},
%Y. S.\ Kivshar, H. S.\ Eisenberg, and Y.\ Silberberg,
IEEE J. Quant. Elect. \textbf{39}, 31 (2003).
 

\bibitem{mplb} V. A. Brazhnyi and V. V. Konotop, Mod. Phys. Lett. B
\textbf{18}, 627 (2004);
O. Morsch and M. Oberthaler,
Rev. Mod. Phys. {\bf 78}, 179 (2006).


\bibitem{IJMPB} P.G. Kevrekidis {\it et al.}, 
%K.{\O}. Rasmussen and A.R. Bishop,
Int. J. Mod. Phys. B {\bf 15}, 2833 (2001).


\bibitem{dnls_book} P.G. Kevrekidis,
{\it The Discrete Nonlinear Schr{\"o}dinger Equation},
Springer-Verlag (Heidelberg, 2009).

\bibitem{SC} B. Grammaticos {\it et al.}, 
Phys. Rev. Lett. {\bf{67}}, 1825  (1991).

\bibitem{Gram} A. Ramani {\it et al.}, 
%{\sl An integrability test for differential-difference systems}, 
J. Phys. A {\bf{25}}, L883 (1992).

\bibitem{comech} A. Comech and D. Pelinovsky,
Comm. Pure Appl. Math {\bf 56}, 1565 (2003).

\bibitem{townes} R.Y. Chiao {\it et al.}, 
%E. Garmire, and C.H. Townes, 
Phys. Rev. Lett. {\bf 13}, 479 (1964).

\bibitem{sivan} Y. Sivan {\it et al.}, 
%G. Fibich, B. Ilan, and M.I. Weinstein,
Phys. Rev. E {\bf 78}, 046602 (2008).

\bibitem{yaron} Y. Silberberg, 
Opt. Lett. {\bf 15}, 1282 (1990).

\bibitem{peli2d} D.E. Pelinovsky {\it et al.}, 
%P.G. Kevrekidis and D.J. Frantzeskakis,
Physica D {\bf 212}, 20 (2005).

\bibitem{bona}  J. Angulo {\it et al.}, 
%J.L. Bona, F. Linares, M. Scialom,
Nonlinearity {\bf 15}, 759 (2002).

\end{thebibliography}
\end{document}